\documentstyle[12pt,epsfig]{article}
\def\bge{\begin{equation}}
\def\ene{\end{equation}}
\def\bg{\begin{eqnarray}}
\def\en{\end{eqnarray}}
\def\nn{\nonumber}

\def\aleq{\stackrel{<}{\sim}}

\def\noi{\noindent}
\begin{document}
\renewcommand{\thefootnote}{\fnsymbol{footnote}}
\begin{flushright}
ADP-97-12/T249 
\end{flushright}
%
%
\begin{center}
\begin{LARGE}
Naturalness in the quark-meson coupling model 
\end{LARGE}
\end{center}
\vspace{0.5cm}
\begin{center}
\begin{large}
K.~Saito\footnote{ksaito@nucl.phys.tohoku.ac.jp} \\
Physics Division, Tohoku College of Pharmacy \\ Sendai 981, Japan \\
K.~Tsushima\footnote{ktsushim@physics.adelaide.edu.au} and 
A.~W.~Thomas\footnote{athomas@physics.adelaide.edu.au} \\
Department of Physics and Mathematical Physics \\
and \\
Special Research Center for the Subatomic Structure of Matter \\
University of Adelaide, South Australia, 5005, Australia
\end{large}
\end{center}
\vspace{0.5cm}
\begin{abstract}
The quark-meson coupling (QMC) model is examined using Georgi's ``naive 
dimensional analysis''.  We argue that the QMC model is quite {\em natural\/} 
as an effective field theory for nuclei. 
\end{abstract}
\vspace{1.5cm}
\newpage

Recently we have developed an effective field theory for both infinite 
nuclear matter and finite nuclei, the quark-meson coupling (QMC) 
model.  (The model originated with Guichon in 1988~\cite{g1}.) 
The QMC model may be viewed as an extension of QHD~\cite{qhd,qhd2} in 
which the nucleons still interact through the exchange of scalar and 
vector mesons.  However, the mesons couple not to point-like nucleons but to
confined quarks (in the bag).  In studies of infinite nuclear 
matter~\cite{g1,st1} 
it was found that the extra degrees of freedom provided by the internal 
structure of the nucleon give an acceptable value for the 
incompressibility once the coupling constants are chosen to 
reproduce the correct saturation energy and density for symmetric 
nuclear matter. This is a significant improvement on QHD at the same level 
of sophistication.  Furthermore, the model has been successfully applied to 
finite nuclei within the Born-Oppenheimer approximation~\cite{st5,blun,st6}. 
It has been found that 
the QMC model can reproduce the properties of finite, closed-shell nuclei 
(from $^{12}$C to $^{208}$Pb) quite well. For this reason it also
provides a very attractive framework within which to develop a
microscopic understanding of the nuclear EMC effect~\cite{st2,jin}.

In general, an effective field theory at low energy will contain an infinite 
number of interaction terms, which incorporate the {\em compositeness\/} of 
the low-energy degrees of freedom, namely the hadrons~\cite{qhd2,qhd3}, and it 
is then expected to involve numerous couplings which may be nonrenormalizable. 
Thus, one needs an {\em organizing principle\/} to make sensible 
calculations.  

Manohar and Georgi~\cite{nda} have proposed a systematic way to 
manage such complicated, effective field theories called ``naive
dimensional analysis'' (NDA).
NDA gives rules for assigning a coefficient of the 
appropriate size to any interaction term in an effective lagrangian.  
(This NDA has been extended to an effective hadronic lagrangian for 
nuclei~\cite{qhd3}.)  After extracting the dimensional factors and 
some appropriate counting factors using NDA, the remaining 
{\em dimensionless\/} 
coefficients are all assumed to be of order {\em unity}.  This is the 
so-called {\em naturalness assumption}.  If naturalness is valid, the 
effective lagrangian can be truncated at a given order with a reasonable 
bound on the truncation error for physical observables.  Then we can control 
the effective lagrangian, at least at the tree level.  

NDA has been 
already applied to QHD~\cite{qhd2,qhd3}, the Nikolaus-Hoch-Madland (NHM) 
model~\cite{nhm,nhm2} and some nuclear phenomena~\cite{some}. 
(Concerning QHD, Furnstahl et al.~\cite{qhd3} concluded that the 
relativistic Hartree approximation (RHA) leads to {\em unnaturally} large 
coefficients due to the treatment of the vacuum in terms of 
$N{\bar N}$-pairs excitation. This means that the loop expansion in QHD 
does not work as well as one would like~\cite{qhd3,prak}.)  

Here we use NDA to see whether the QMC model gives natural coefficients.
In brief, NDA tells us the following: for the strong interaction 
there are two relevant 
scales, namely, the pion-decay constant $f_\pi$ ($\sim$ 93 MeV) and a larger 
scale, $\Lambda \sim$ 1 GeV, which characterizes the mass scale of physics 
beyond the Goldstone bosons.  The NDA rules indicate how those scales should 
appear in a given term in the effective lagrangian.  The rules are: 
\begin{enumerate}
\item include a factor of $ 1/f_\pi $ for each strongly interacting field, 
\item assign an overall normalization factor of $(f_\pi \Lambda)^2$, 
\item multiply by factors of $1/\Lambda$ to achieve dimension (mass)$^4$, 
\item include appropriate counting factors, e.g. 1/$n!$ for $\phi^n$ 
(where $\phi$ 
is a meson field). 
\end{enumerate}
Since the QMC lagrangian in the mean-field approximation (MFA) is given in 
terms of the nucleon ($\psi$), scalar 
($\sigma$) and vector ($\omega$ and $\rho$) meson fields, we can scale a 
generic lagrangian component as 
\bge
{\cal L} \sim c_{\ell m n p} \frac{1}{m! n! p!} 
\left( \frac{{\bar \psi} \Gamma (\tau/2) \psi}{f_\pi^2 \Lambda} \right)^{\ell} 
\left( \frac{\sigma}{f_\pi} \right)^m 
\left( \frac{\omega}{f_\pi} \right)^n 
\left( \frac{b}{f_\pi} \right)^p (f_\pi \Lambda)^2 , 
\label{general}
\ene
where $\Gamma$ and $\tau$ stand for a combination of Dirac matrices and 
isospin operators.  The $\rho$ meson field is denoted by $b$. 
The overall coupling constant $c_{\ell m n p}$ is 
dimensionless and of ${\cal O}(1)$ if naturalness holds.  

As shown in Ref.~\cite{st5}, the basic result in the QMC model with MFA 
is that, in the scalar and vector meson 
fields, the nucleon behaves essentially as a point-like particle with an 
effective mass $m_N^{\star}$, which is given by a (relativistic) quark model 
for the 
nucleon (like a bag) and depends on the nuclear density through only the 
$\sigma$ field, moving in a vector potential generated by the vector mesons.  
Because of their Lorentz-vector character, in infinite nuclear matter 
the vector mesons provide {\em no effect 
on the nucleon structure} except for an overall phase in the quark wave 
function, which gives a shift in the nucleon energy.  
In an earlier version of the QMC 
model~\cite{g1,st1,st5,blun}, we considered the effect of 
the nuclear medium on the structure of the nucleon alone 
and froze the quark degrees of freedom 
in the mesons.  We call this version QMC-I.  

However, it is true that the mesons are also built of quarks and anti-quarks, 
and that they may change their character in matter~\cite{st6,st6p}.  
To incorporate the effect of meson structure in the QMC model in MFA, we 
suppose that the vector mesons are again described by a relativistic quark 
model with {\em common\/} scalar and vector mean-fields, 
like the nucleon in QMC-I.  In this case the effective vector-meson mass in 
matter, $m_v^{\star} (v = \omega, \rho)$, will also depend on the $\sigma$ 
mean-field.  

The $\sigma$ meson itself is, however, not so readily represented  
by a simple quark model (like a bag), because it couples strongly 
to the pseudoscalar ($2 \pi$) channel and a direct 
treatment of chiral symmetry in medium is important~\cite{hatkun}.  
On the other hand, many approaches, including 
the Nambu--Jona-Lasinio model~\cite{hatkun}, the Walecka 
model~\cite{sai} and Brown-Rho scaling~\cite{brown} suggest that the 
$\sigma$-meson mass in medium, $m_{\sigma}^{\star}$, should   
be less than the free 
one, $m_\sigma$.  We therefore parametrized it using a quadratic 
function of the scalar field (by hand)~\cite{st6}: 
\bge
\left( \frac{m_{\sigma}^{\star}}{m_{\sigma}} \right) = 1 - a_{\sigma} 
(g_{\sigma} \sigma) + b_{\sigma} (g_{\sigma} \sigma)^2 , 
\label{sigmas}
\ene
with $g_\sigma \sigma$ in MeV, where $g_\sigma$ is the $\sigma$-nucleon 
coupling constant (in free space).  
In Ref.~\cite{st6}, we chose three parameter sets: 
($a_\sigma$ ; $b_\sigma$) = (3.0, 5.0 and 7.5 $\times 10^{-4}$ MeV$^{-1}$ ; 
10, 5 and 10 $\times 10^{-7}$ MeV$^{-2}$), called A, B and C, respectively.
The values for the sets A, B and C were determined so as 
to reduce the mass of the $\sigma$-meson 
by about 2\%, 7\% and 10\% (respectively)  
at saturation density.  This version, involving the structure 
effects of both the nucleon and the mesons, was called QMC-II.  

The QMC lagrangian in MFA is given by~\cite{st5,st6} 
\bge
{\cal L}_{QMC} = {\cal L}_{free} + {\cal L}_{em} + {\cal L}_{QMC}^{int.} , 
\label{QMCgen}
\ene
where ${\cal L}_{free}$ and ${\cal L}_{em}$ 
stand for the free lagrangian for the nucleon and 
mesons and the electromagnetic interaction, respectively, while 
${\cal L}_{QMC}^{int.}$ involves (strong) interaction terms.  
For QMC-I and QMC-II, ${\cal L}_{QMC}^{int.}$ is respectively given 
by~\cite{st6} 
\bge
{\cal L}_{QMC-I}^{int.} = {\bar \psi} \left[ 
\left(m_N - m_N^{\star}(\sigma) \right) 
- g_\omega \gamma_0 \omega - g_\rho \left( \frac{\tau_3^N}{2} \right) 
\gamma_0 b \right] \psi, 
\label{QMC-I}
\ene
and 
\bg
{\cal L}_{QMC-II}^{int.} &=& {\bar \psi} \left[ 
\left(m_N - m_N^{\star}(\sigma) \right) 
- g_\omega \gamma_0 \omega - g_\rho \left( \frac{\tau_3^N}{2} \right) 
\gamma_0 b \right] \psi \nn \\
&-& \frac{1}{2} \left[ m_\sigma^{\star 2}(\sigma)-m_\sigma^2 \right] \sigma^2 
+ \frac{1}{2} \left[ m_\omega^{\star 2}(\sigma)-m_\omega^2 \right] \omega^2 \\
&+& \frac{1}{2} \left[ m_\rho^{\star 2}(\sigma)-m_\rho^2 \right] b^2 , \nn 
\label{QMC-II}
\en
where $g_\omega$ and $g_\rho$ are 
the $\omega$- and $\rho$-nucleon coupling constants, respectively. 
The effective masses in the medium,
$m_j^{\star}$ ($j = N, \sigma, \omega, \rho$), 
depend only on the scalar field, and $m_j$ is the free 
mass.  If we introduce the field-dependent $\sigma$-nucleon 
coupling constant by 
\bge
m_N^{\star} \equiv m_N - g_\sigma(\sigma) \sigma, 
\label{efcc}
\ene
the nucleon structure effect can be 
completely absorbed into the coupling constant $g_\sigma(\sigma)$.  
(Note that $g_\sigma(0) = g_\sigma$ in (\ref{sigmas}).) 

To determine the $\sigma$-dependence of $m_j^{\star}$, we need a model for the 
structure of the hadron involved.  For simplicity, we use the MIT bag model. 
(An alternative model can be found in Ref.~\cite{blun}.)  
Then the mass at nuclear density $\rho_B$ is found to take quite a simple 
form (for $\rho_B \aleq 3 \rho_0$, where $\rho_0$ is the 
saturation density for normal nuclear matter)~\cite{st6}: 
\bge
m_j^{\star} \simeq m_j - \frac{n_0}{3} (g_\sigma \sigma) \left[ 1 - 
  \frac{a_j}{2} (g_\sigma \sigma) \right] , 
\label{efmas}
\ene
where $n_0$ is the number of non-strange quarks in the hadron $j$ and 
$a_j$ is a slope parameter for the hadron $j$, which is given by the 
second derivative of the mass with respect to the $\sigma$ field. 

There are some coupling constants to be determined: 
the $\sigma$- and $\omega$-nucleon coupling constants ($g_\sigma$ and 
$g_\omega$) are fixed by fitting the 
binding energy ($-15.7$ MeV) at the correct saturation density 
($\rho_0 = 0.15$ fm$^{-3}$) for symmetric nuclear matter.  The 
$\rho$-nucleon coupling constant $g_\rho$ is used to reproduce the bulk 
symmetry energy, 35 MeV.  The slope parameters and the coupling constants 
are listed in Table~\ref{cc}. 
(In the actual calculation we take the bare quark mass and 
the bag radius of the free nucleon to be 5 MeV and 0.8 fm, respectively, and 
$m_N$ = 939 MeV, $m_\sigma$ = 550 MeV, 
$m_\omega$ = 783 MeV and $m_\rho$ = 770 MeV.  For details, see 
Ref.~\cite{st6}.) 
\begin{table}[htbp]
\begin{center}
\caption{Coupling constants (dimensionless)
and slope parameters (in units of $10^{-4}$ 
MeV$^{-1}$) for QMC-I and QMC-II (sets A -- C).  }
\label{cc}
\begin{tabular}[t]{c|cccccc}
\hline
type & $g_\sigma$ & $g_\omega$ & $g_\rho$ & $a_N$ & $a_\omega$ & $a_\rho$ \\
\hline
 QMC-I & 8.24 & 8.17 & 9.33 & 8.80 & --- & --- \\
 A &  6.95 & 5.82 & 8.35 & 9.01 & 8.63 & 8.59 \\
 B &  7.04 & 6.31 & 8.14 & 8.98 & 8.63 & 8.58 \\
 C &  6.94 & 6.45 & 8.07 & 8.97 & 8.63 & 8.58 \\
\hline
\end{tabular}
\end{center}
\end{table}

Using (\ref{sigmas}) and (\ref{efmas}), the QMC-I lagrangian gives four 
interaction terms, while the QMC-II lagrangian offers 16 terms 
due to the internal structure of the nucleon {\em and\/} the mesons.  
Now we can calculate the dimensionless 
coefficients, $c_{\ell m n p}$ in (\ref{general}), for each 
interaction term using NDA.  Table~\ref{coef} summarizes the interaction 
terms and the corresponding dimensionless coefficients. 
(We take $\Lambda = m_N$ in (\ref{general}).) 
\begin{table}[htbp]
\begin{center}
\caption{Interaction terms and corresponding (dimensionless) coupling 
constants. }
\label{coef}
\begin{tabular}[t]{c|c|cccc}
\hline
term & $c_{\ell m n p}$ & QMC-I & A & B & C \\
\hline
${\bar \psi} \sigma \psi$ & $c_{1100}$ & 0.82 & 0.69 & 0.70 & 0.69 \\
${\bar \psi} \sigma^2 \psi$ & $c_{1200}$ & -0.55 & -0.40 & -0.41 & -0.40 \\
${\bar \psi} \gamma_0 \omega \psi$ & $c_{1010}$ & -0.81 & 
                                                 -0.58 & -0.63 & -0.64 \\
${\bar \psi} \left( \frac{\tau}{2} \right) \gamma_0 b \psi$ & $c_{1001}$ & 
                                         -0.92 & -0.83 & -0.81 & -0.80 \\
$ \sigma^3 $ & $c_{0300}$ & --- & 0.40 & 0.67 & 1.0 \\
$ \sigma^4 $ & $c_{0400}$ & --- & -3.6 & -2.2 & -4.4 \\
$ \sigma^5 $ & $c_{0500}$ & --- & 3.3 & 2.9 & 8.3 \\
$ \sigma^6 $ & $c_{0600}$ & --- & -22 & -5.7 & -22 \\
$ \sigma \omega^2 $ & $c_{0120}$ & --- & -0.77 & -0.78 & -0.76 \\
$ \sigma^2 \omega^2 $ & $c_{0220}$ & --- & 0.85 & 0.87 & 0.85 \\
$ \sigma^3 \omega^2 $ & $c_{0320}$ & --- & -0.71 & -0.73 & -0.70 \\
$ \sigma^4 \omega^2 $ & $c_{0420}$ & --- & 0.39 & 0.41 & 0.39 \\
$ \sigma b^2 $ & $c_{0102}$ & --- & -0.75 & -0.76 & -0.75 \\
$ \sigma^2 b^2 $ & $c_{0202}$ & --- & 0.84 & 0.86 & 0.84 \\
$ \sigma^3 b^2 $ & $c_{0302}$ & --- & -0.70 & -0.73 & -0.70 \\
$ \sigma^4 b^2 $ & $c_{0402}$ & --- & 0.39 & 0.41 & 0.39 \\
\hline
\end{tabular}
\end{center}
\end{table}

As seen in the table, the QMC-I model provides remarkably {\em natural\/} 
coupling constants, which lie in the range 0.5 -- 1.0.  
In QMC-II, 14 or 15 of the 16 coupling constants can be regarded as 
{\em natural}.  Only the large value of
$c_{0500}$ for set C and $c_{0600}$ for 
sets A -- C are unnatural. 
Since the coefficients, $c_{0500}$ and $c_{0600}$, are respectively 
proportional to $a_\sigma b_\sigma$ and $b_\sigma^2$ ($a_\sigma$ and 
$b_\sigma$ are defined by (\ref{sigmas})), we can see that those 
unnaturally large numbers are due to the parametrization of the $\sigma$ mass 
in matter.  In particular, a large value for the coefficient  
$b_\sigma$ leads to unnaturally values for 
$c_{0500}$ and $c_{0600}$.  (This is the reason why $c_{0500}$ and 
$c_{0600}$ in set B are much closer to natural than in sets A and C.  
See also below (\ref{sigmas}).)  
We recall that the reduction of the $\sigma$-meson mass in matter was
the sole feature of the model which could not be calculated but was put
in by hand. There is nothing within the QMC model which requires
$b_\sigma$ to be so large.
Therefore, the QMC model itself can be regarded as a 
{\em natural\/} effective field theory for nuclei. 

Finally, we compare the QMC model with the NHM model~\cite{nhm} and a new 
lagrangian, which was constructed recently by Furnstahl, Serot, 
Tang and Walecka~\cite{qhd2,qhd3} 
(we call the latter the FSTW model).  The NHM model was 
motivated by empirically based improvements to QHD~\cite{qhd}, but 
using contact (zero-range) interactions to allow treatment of the Fock terms. 
It has 9 coupling constants, and 6 of them are 
{\em natural}~\cite{nhm2}.  On the other hand, the FSTW model was constructed 
in terms of nucleons, pions and the low-lying non-Goldstone bosons, and chiral 
symmetry is realized nonlinearly with a light scalar ($\sigma$) meson 
included as a chiral singlet to describe the mid-range nucleon-nucleon 
attraction.  This model has a total of 16 coupling constants, 
and they are almost all {\em natural}.  
In both cases, the coupling 
constants were determined so as to fit measured ground-state observables 
of several nuclei by a self-consistent procedure that 
solved the model equations for the nuclei simultaneously and minimized 
the difference between the measured and calculated quantities using a 
nonlinear least-squares adjustment algorithm.  Therefore, the coupling 
constants were fixed entirely phenomenologically.  

This is quite different from the case of the QMC model.  
The latter has basically three coupling 
constants, which were determined to fit the saturation properties of 
nuclear matter (as mentioned after (\ref{efmas})), but the other coupling 
constants are automatically given through a model for the structure of the 
hadrons.  Therefore, the physical meaning of the coupling constants is quite 
clear.  In the QMC model, the meson mass ($\sigma$, $\omega$, $\rho$ etc) 
decreases in matter.  However, in the FSTW model some parameter sets lead to 
an increase of the effective meson masses in matter, which seems unlikely 
from the point of view of the recent discussions on this matter by many 
authors~\cite{hatkun,sai,brown,hat,lein}.  

In conclusion, we have demonstrated that the QMC model 
is quite {\em natural\/} as 
an effective field theory for nuclei. 
Although we discussed the QMC model using 
a specific hadronic model, namely the MIT bag, 
the qualitative features we found here are expected to hold 
in any model in which the 
hadrons contain relativistic quarks.  

\vspace{1cm}
\noi This work was supported by the Australian Research Council.  
%
%

%
\end{document}